\documentclass[english]{article}
\usepackage{babel, mathrsfs, bbm, amsmath, amssymb}
\usepackage{bm}
\usepackage[T1]{fontenc}
\usepackage[latin1]{inputenc}

\begin{document}

\title{Definition of a time variable with Entropy of a perfect fluid in Canonical Quantum Gravity}

\author{
Francesco Cianfrani$^{\dag}$, Giovanni Montani$^{\dag\ddag*}$, Simone Zonetti$^\S$ \\
{\small $^\dag$Dipartimento di Fisica, Universit\`a degli Studi di Roma ``Sapienza'',}\\
{\small Piazzale Aldo Moro 5, 00185 Rome, Italy,} \\
{\small $^\ddag$ENEA C.R. Frascati (Dipartimento F.P.N.), via Enrico Fermi 45,}\\ 
{\small 00044 Frascati, Rome, Italy,}\\
{\small $^*$ ICRANET C. C. Pescara, Piazzale della Repubblica, 10, 65100 Pescara, Italy}\\
{\small $^\S$ Center for Particle Physics and Phenomenology (CP3)}\\ 
{\small Universit\`e Catholique de Louvain, Chemin du Cyclotron, 2}\\ 
{\small B-1348 Louvain-la-Neuve, Belgium}
}

\date{April 2009}

\maketitle

\begin{abstract}
The Brown-Kucha\v r mechanism is applied in the case of General Relativity coupled with the Schutz' model for a perfect fluid. Using the canonical formalism and manipulating the set of modified constraints one is able to recover the definition of a time evolution operator, {\it i.e.} a physical Hamiltonian, expressed as a functional of gravitational variables and the entropy.
\end{abstract}

\section{Introduction}

The problem of time \cite{Isham:1992a} is a well known issue in the context of the canonical approaches to Quantum Gravity \cite{rovelli:2004} \cite{rovelli:1997}. In General Relativity (GR) the time dimension acquires a new meaning: it is no more an external parameter, as in classical Newtonian and quantum mechanics, but it is considered as part of a dynamical object, the metric field \cite{wald:1984}. This has far reaching consequences in the search for a quantum theory of gravity, above all in the canonical approach.\\
If one tries to follow the standard canonical quantization procedure \cite{hanneaux:1994} \cite{dirac:1964} the problem soon arises: the most important outcome of such procedure is that \emph{the Hamiltonian of GR is a linear combination of constraints}, the so called super-Hamiltonian $H^G$ and super-momentum $H^G_i$:
\begin{equation}
\mathcal{H} = \int d^3 x (H^G N + H^G_a N^a) \approx 0,
\end{equation}
so \emph{it vanishes}. Considering the Hamiltonian operator $\hat{\mathcal{H}}$ as the generator of time evolution in the quantum theory, one obtains the well known Wheeler-DeWitt equation \cite{dewitt:1967a} \cite{dewitt:1967b}:
\begin{equation}
\hat{\mathcal{H}} \psi = 0,
\end{equation}
and easily sees that there is no possibility for an evolutionary picture.\\
Different approaches have been developed, trying to recover a meaningful definition of time \cite{Isham:1992a}  \cite{ashtekar:1987} \cite{montani:2007} \cite{cianfrani:2008} \cite{cianfrani:2007}: a possible solution is to couple matter to the gravitational field. It can be shown that there is a strong duality between matter and reference frames (\cite{thiemann:2006} \cite{cianfrani:2008} \cite{mercuri:2003}, see \cite{kuchartorre:1991.1} and \cite{kuchartorre:1991.2} for a clear example), as one can intuitively use fluid particles to label space-time points, while general covariance can be recovered simply re-parameterizing the coordinate dependence.\\
In fact adding coordinate conditions (enforced with Lagrange multipliers) to the action of GR gives new terms in the re-parametrized action, which can be seen, in the Einstein equations, as the matter terms. Viceversa matter terms can be seen as coordinate conditions as well \cite{Isham:1992a}.\\
This duality can be used in a more convenient way through the Brown-Kucha\v r mechanism \cite{kucharbrown:1995}, which, by manipulating the set of constraints of General Relativity coupled with matter, enables one to recover the definition of a meaningful, non vanishing, evolution operator.\\
In this paper this procedure is applied to the Schutz' model of a perfect fluid \cite{schutz:1971}. This is a 
much more realistic description of the cosmological fluid with respect to the dust used in \cite{kucharbrown:1995}, 
especially when the cosmological singularity is approached. Indeed the primordial thermal bath is not characterized 
by a dust equation of state ($p=0$), but it is better described by an ultra-relativistic perfect fluid (having 
equation of state $p=\epsilon/3$) since the thermal energy is much greater than the mass of the particles living in
 the equilibrium. Despite the Schultz fluid is not intrinsically an ultra-relativistic object, it provides a proper 
representation of a fluid constituted by particles whose own entropy is conserved. This scheme is very suitable to 
describe the initial phases of the Universe evolution, especially when we deal with a Universe as an open 
thermodynamical system. Recent studies in cosmology outlined the possibility that the Universe is characterized by 
a non-trivial evolution of the equation of state, as for instance it is predicted in late time to describe the so-called Dark 
Energy phenomenon \cite{revDE}. Such a feature is properly realized when a Schutz' fluid is addressed, and the equation of state 
will depend on the specific thermodynamical condition of the fluid.

Thus any reliable implementation to the cosmological framework of a Quantum Gravity theory based on the Kucha\v r-Brown 
formulation requires an extension of the original work to make account of interaction features among the 
fundamental constituents of the Universe. In particular, we outline that the Brown-Kucha\v r mechanism 
works also in this case and it gives a strong temporal value to the entropy field associated with this particular 
matter field. This is an intriguing result in view of finding a connection between the thermodynamical time and 
time in Quantum Gravity.\\
This work consists in five more sections.
: in section 2 the Brown-Kucha\v r mechanism will be reviewed in the case of a generic scalar field Lagrangian, while in section 3 the Schutz' fluid will be described in the thermodynamical context. In section 4 the Hamiltonian formulation for the uncoupled model will be presented, and in section 5 the mechanism will be finally applied to the coupled model. In section 6 results and perspectives are spotted out.\\
\section{The Brown-Kucha\v r Mechanism}\label{thekbmechanism}
We review the Brown-Kucha\v r mechanism for a generic scalar field Lagrangian, whose form is the following one
\begin{equation}
\mathcal{L}_F = \mathcal{L}_F [-\phi_{,\mu}\phi^{,\mu}] = \mathcal{L}_F [\Upsilon],
\end{equation}
where the comma denotes the ordinary derivative and indices are contracted with the space-time metric.\\
To keep the general covariance of the theory untouched and to identify the formal time variable needed in the canonical formalism, one performs a standard 3+1 ADM splitting of the space-time \cite{ADM:1962} \cite{thiemann:2001}, with the assumption $\mathcal{M} = \mathbbm{R} \times \sigma$ on the 4-dim space-time manifold $\mathcal{M}$, that gives, dropping commas denoting ordinary derivatives of the scalar field without ambiguities:
\begin{subequations}\label{splitting}
\begin{align}
g_{\mu \nu} &= -n_\mu n_\nu + q_{\mu \nu} \\
\begin{split}
\dot \phi &= \mathscr{L}_T \phi = \phi_{\mu} T^{\mu} =  \\
&=\phi_{\mu} n^{\mu} N + \phi_a N^a = \phi_n N + \phi_a N^a, \label{fipunto}
\end{split}\\
-\phi_\mu \phi^\mu &= (\phi_\mu n^{\mu})^2 - \phi_a \phi^a = (\phi_n )^2 - V, \label{demufi}
\end{align}
\end{subequations}
where $N$ and $N_a$ are the lapse function and the shift vector respectively, $n^\mu$ is the normal to the 3-dim hypersurfaces and $T^\mu$ is the deformation vector.
Greek indices run from 0 to 3, labeling the 4-dim manifold coordinates, latin indices run from 1 to 3 and label coordinates on the 3-dim manifold.\\
The calculation of the conjugate momentum is now straightforward:
\begin{equation}
\pi = \frac{\delta \mathcal{S}_F}{\delta \dot{\phi}} = -2\sqrt{q} \phi_n \frac{\delta \mathcal{L}_F[\Upsilon]}{\delta \Upsilon},
\end{equation}
where one has defined $q = det(q_{ab})$, which can be seen as an equation for $\Upsilon$.
It can be solved together with the splitting of $\Upsilon$ given by \eqref{demufi}, and the system will have solutions $\Upsilon = \tilde{\Upsilon}[\pi, V]$ and 
\begin{equation}
\phi_n =\Phi[\pi, V]= \frac{\pi}{2 \sqrt{q}}\left[ \frac{\delta \mathcal{L}_F}{\delta \Upsilon} \right]^{-1}.
\end{equation}
This allows to complete the Legendre transformation, so that the Hamiltonian will contain only spatial quantities and the conjugate momentum $\pi$, and will take the simple form:
\begin{equation}
\mathcal{H}_F = \int d^3 \! x (H^F N + H^F_a N^a),
\end{equation}
where $H^F$ is a suitable defined functional of $\pi$, $V$ and the 3-dim metric $q_{ab}$, and $H^F_a = \phi_a \pi$.\\
By adding the Einstein-Hilbert action of GR the Hamiltonian density will simply be the sum of the matter-free super-Hamiltonian and super-momentum with the scalar field terms defined above, because there are no derivatives of the metric tensor $g_{\mu \nu}$ in the field action.
The modified constraints of GR will read
\begin{subequations}
\begin{align}
H = & H^G[q,P] + H^F[\pi, V, q],\\
H_a = & H^G_a [q,P] + \phi_a \pi.
\end{align}
\end{subequations}
At this point one can square the super-momentum and solve it for $V$. \emph{This allows one to completely eliminate the spatial gradients from the super-Hamiltonian}, which can be seen as an equation for $\pi$. Assuming that in turn $H$ could be solved for it, one obtains an equivalent constraint; in general there will be multiple solutions, which can be selected using consistency conditions with the properties of the model, like the closed Dirac algebra of secondary constraints of GR. Upon resolution it takes the form:
\begin{equation}\label{eqconstraint}
\pi - h[ q, P] = 0.
\end{equation}
This is the starting point to the construction of a physical Hamiltonian, since, as soon as the variable $\phi$ is taken as internal time and 
the canonical quantization of this system is performed, equation \eqref{eqconstraint} can be expected to lead to a Schr\" odinger equation. The physical Hamiltonian will simply consist in $\mathcal{H}_{phys} = \int d^3 \! x h(x)$, but it will need to fulfill some necessary conditions:
\begin{itemize}
\item Independence from the field whose conjugate momentum is $\pi$.
\item Invariance under the 3-diff group, i.e. under the action of the super-momentum, considered as the generator of the spatial diffeomorphisms. This is manifest if $h$ is a scalar density of weight one.
\item Invariance under re-parametrization of the time-like variable, {\it i.e.} $\{h,h\}=0$ strongly \cite{kucharbrown:1995}:
it tell us that 4-diff invariance is not broken explicitly. In the case of multiple solutions this condition may be used to eliminate the unphysical ones.
\end{itemize}
Once these properties are checked, one is able to write the evolution equation for observables as the action through the Poisson brackets of $\mathcal{H}_{phys}$:
\begin{equation}
-\frac{d\mathcal{O}(t)}{dt } = \{ \mathcal{H}_{phys}, \mathcal{O}(t) \}.
\end{equation}
Here one assumes that observables exist, and have the standard property of vanishing Poisson brackets with the whole set of constraints.

\section{Standard description of baryonic fluid}

\subsection{Thermodynamics}
One considers a perfect fluid composed by baryons, whose true mass may not be conserved, but the baryon number $N$ is. So one can define the conserved rest mass as $m_HN=\rho_0$, where $m_H$ is the mass of an hydrogen atom in its ground state.
The specific internal energy $\Pi=U/\rho_0$ will account for the difference with the real rest mass of the system, as well for electron-positron pairs, photons, thermal motion and so on.\\
The density of total mass energy is then $\rho= \rho_0 (1+\Pi)$.\\
Following E. Fermi \cite{fermi:1936} a two parameter equation of state is associable with this one component fluid, so $p=p(\rho_0, \Pi)$.\\
From the first law of thermodynamics one can compute the amount of energy per unit rest mass added to the fluid in a quasistatic process:
\begin{equation}
\delta q = d\Pi + pd(1/ \rho_0 ),
\end{equation}
and using the Pfaff's theorem one can claim that there exist functions $S(\rho_0, \Pi)$, specific entropy, and $T(\rho_0, \Pi)$, temperature, such that:
\begin{equation}
d\Pi + pd(1/\rho_0) = TdS= \delta q.
\end{equation}
Defining now the specific inertial mass as $\mu = (\rho + p)/\rho_0 = p/\rho_0 + \Pi + 1$, one can eliminate $d\Pi$ in the last equation:
\begin{equation}
d\Pi = -d(p/\rho_0) + d\mu
\end{equation}
and obtain, by integration, the equation of state:
\begin{equation}\label{eqdistato}
p=\rho_0 ( \mu - TS ).
\end{equation}

In terms of the energy density $\rho$, the equation above can be written as
\begin{equation}
p=\left(\frac{\mu}{1+\Pi}-1\right)\rho,
\end{equation}
thus the equation of state contains a dependence on internal degrees of freedom of the fluid and in general it is expected to vary in time. Hence the whole dynamical system is much more complicated than the standard dust fluid and, in a cosmological setting, we think it better approximates the behavior of the thermal bath.

\subsection{Stress-Energy Tensor and Equations of Motion}
The associated stress-energy tensor has the standard form:
\begin{equation}\label{stressenergyt}
T^{\mu \nu} = (p + \rho)U^\mu U^\nu + p g^{\mu \nu}.
\end{equation}
The conservation of the baryon number $N$ is contained in the equation:
\begin{equation}\label{baryonN}
(\rho_0 U^\nu)_{;\nu}=0,
\end{equation}
while the normalization of the 4-velocity $U^\mu$ leads to the condition:
\begin{equation}\label{normaliz}
U^\nu U_{\nu ;\mu}=0.
\end{equation}
Finally, the equations of evolution can be obtained by requiring the stress-energy tensor to be divergence free:
\begin{equation}
T^{\mu \nu}_{\phantom{\mu \nu} ;\mu}=0.
\end{equation}
Looking for the components perpendicular to $U^\mu$ and using eqs. (\ref{baryonN}) and (\ref{normaliz}) one gets:
\begin{equation}
\rho_0 T U^\nu S_{,\nu}=0,
\end{equation}
that states that the entropy per baryon is conserved in the motion of the perfect fluid.

\section{Velocity-Potentials representation}
\subsection{Variational principle in hydrodynamics}
The Hamiltonian formulation of a generally covariant theory of hydrodynamics requires a standard variational principle to be applied to the action, so it is necessary to develop a field-like formalism able to describe fluids in this frame. 
It is well known that irrotational motion can be derived from a single potential $\boldsymbol{v}= \nabla \phi$, and a result established by Clebsch \cite{clebsch:1859} in 1859 claims that any motion can be represented by three potentials:
\begin{equation}\label{3potentials}
\boldsymbol{v}=\nabla \phi + \alpha \nabla \beta.
\end{equation}
This representation has one great disadvantage: the potentials have not a direct physical meaning. This is because there are no individual evolution (in the Eulerian sense) equations that give the variation of $\boldsymbol{v}$ without the imposition of external conditions; and these conditions come directly from to the hydrodynamics equations. So a three-potential model cannot acquire the title of a concrete alternative to standard hydrodynamics, being actually built on it.\\
Therefore an extension of this model is required: Seliger and Whitham \cite{seligerwhitham:1968} propose a five-potentials model for non-relativistic fluid, where each potential has an equation of evolution and one of them is interpreted as the entropy. A relativistic generalization requires an additional potential \cite{schutz:1970} \cite{schutz:1971}. This 6-potentials model is self-consistent, and does not need any condition in order to reproduce the usual equations of hydordynamics.

\subsection{Fluid action}
A general formulation with velocity potentials is based on the fact that the 4-velocity $U^\mu$ of the fluid can be expressed as a combination of scalar fields and their gradients:
\begin{equation}\label{4velocity}
U_\nu = \mu^{-1}(\phi_{,\nu} + \alpha \beta_{,\nu} + \theta S_{,\nu}) = \frac{v_\nu}{\mu},
\end{equation}
where $\mu$ and $S$ are defined as in the previous section. Once more the commas denoting fluid gradients will be dropped with no ambiguities.\\
The specific inertial mass can be fixed by the normalization condition on the 4-velocity, so that $\mu^2=-v^\mu v_\mu$, where the metric is assumed to have signature $(+,-,-,-)$.\\
In this way the equation of state reads, with T fixed:
\begin{equation}\label{equationofstate}
p=\rho_0 ( \sqrt{-v^\mu v_\mu}-TS ),
\end{equation}
where the sign ambiguity that comes from the square root is solved by the request for $\mu$ to be positive and $v^\mu$ to be time-like.\\
Using the stress-energy tensor (\ref{stressenergyt}) and the Einstein equations for matter
\begin{equation}
\delta \mathcal{L}_F /\delta g_{\mu \nu} = T^{\mu \nu}/2,
\end{equation}
one can compute the Lagrangian density for the fluid. With such an integration one can identify:
\begin{equation}\label{saction}
\mathcal{L}_F=\sqrt{-g}p = \sqrt{-g}\rho_0 ( \sqrt{-v^\mu v_\mu}-TS ).
\end{equation}
and so $\mathcal{S}_F=\int d^4 \!x \mathcal{L}_F(x)$.

\subsection{Hamiltonian formulation}
The fluid action so takes the form:
\begin{equation}\label{schutzADMaction}
\mathcal{S}_F=\int_\mathbbm{R} \! dt \int_\sigma \! d^3 \ x  \sqrt{q} N \rho_0 \left( \sqrt{(v_\mu n^\mu)^2 - v_a v^a} - TS \right).
\end{equation}

From the definition of momenta the following constraints are inferred 
\begin{equation}
\chi_1=p_{\alpha} = 0, \qquad
\chi_2=p_{\beta}-  \alpha \pi=0, \qquad
\chi_3=p_{\theta} = 0, \qquad
\chi_4=p_{S}- \theta \pi.\label{sconstraint}
\end{equation}

The Poisson brackets among constraints read as follows 
\begin{equation}
[\chi_A(x),\chi_B(y)]=\left(\begin{array}{cccc} 0 & \pi(y)\delta^3(x-y) & 0 & 0 \\
-\pi(x)\delta^3(x-y) & 0 & 0 & 0 \\ 0 & 0 & 0 & \pi(y)\delta^3(x-y) \\ 0 & 0 & -\pi(x)\delta^3(x-y) & 0  
\end{array}\right),
\end{equation}

thus these constraints are second-class. Let us now restrict to the constraints hypersurfaces parametrized by $X^A=(\phi,\pi,\alpha,\beta,\theta,S)$,  where the induced symplectic form is given by 
\begin{eqnarray*}
\sigma^{AB}(x,y)=\hspace{13cm}\\=\left(\begin{array}{cccccc} 0 & \delta^3(x-y) & -\frac{\alpha(y)}{\pi(y)}\delta^3(x-y) & 0 & -\frac{\theta(y)}{\pi(y)}\delta^3(x-y) & 0 \\\\ -\delta^3(x-y) & 0 & 0 & 0 & 0 & 0 \\\\ \frac{\alpha(x)}{\pi(x)}\delta^3(x-y) & 0 & 0 & -\frac{1}{\pi(x)}\delta^3(x-y) & 0 & 0 \\\\ 0 & 0 & \frac{1}{\pi(y)}\delta^3(x-y) & 0 & 0 & 0 \\\\ \frac{\theta(x)}{\pi(x)}\delta^3(x-y) & 0 & 0 & 0 & 0 & -\frac{1}{\pi(x)}\delta^3(x-y) \\\\ 0 & 0 & 0 & 0 & \frac{1}{\pi(y)}\delta^3(x-y) & 0 \end{array}\right).
\end{eqnarray*}

The Legendre transformation will lead to the formal Hamiltonian function:
\begin{equation}
H=N\left(\sqrt{(\pi^2-q\rho_0^2)V}+q\rho_0TS\right)+N^a\pi v_a,
\end{equation}

where $V = v_a v^a$ and non-spatial quantities have been eliminated using relations below
\begin{subequations}\label{sinverserelations}
\begin{align}
v_\mu v^\mu &= \frac{q \rho_0^2}{\pi^2 - q \rho_0^2} V,\\
(v_\mu n^\mu)^2 &= \frac{\pi^2}{\pi^2 - q \rho_0^2} V.
\end{align}
\end{subequations}

\section{Coupling with General Relativity}
By coupling the fluid with GR one expects the primary constraints to be simply the union of the two sets of primary constraints, since the fluid action does not contain any derivative of the metric tensor $g$. As for the matter-free theory the secondary constraints will be identified with the functionals that appear enforced by the lapse function and the shift vector:
\begin{equation}
\mathcal{H} = \int d^3 \! x (H N + H_a N^a),
\end{equation}
which are
\begin{eqnarray}\label{sgrmultipleH}
H = \sqrt{V(\pi^2 - q \rho_0^2)}+ \sqrt{q} \rho_0 S T + H^G,\\
H_a = \pi v_a + H^G_a.\hspace{2cm}
\end{eqnarray}

The algebra of these constraints is not modified by the presence of the Schultz fluid. In particular, if we introduce smeared functions $f,g$ and $f^a,g^a$, the Poisson brackets between smeared quantities $H[f]=\int_\Sigma d^3x fH$ and $\vec{H}[\vec{f}]=\int_\Sigma d^3x f^aH_a$ are given by the following expressions 
\begin{eqnarray} 
\{\vec{H}[\vec{f}];\vec{H}[\vec{g}]\}=\vec{H}[{\cal L}_{\vec{f}} \vec{g}]\label{alcons1}\\
\{\vec{H}[\vec{f}];H[g]\}=\vec{H}[{\cal L}_{\vec{f}}g]\label{alcons2}\\
\{H[f];H[g]\}=H[\vec{N}(f;g;h)],\label{alcons3} 
\end{eqnarray} 
${\cal L}$ being the Lie derivative, while the expression of the function $\vec{N}(f;f';h)$ is the following
\begin{eqnarray}
N^i(f;f';h)=h^{ij}(f\partial_jf'-f'\partial_jf).
\end{eqnarray}

Therefore, \emph{the full covariance of model is preserved}. Moreover the only secondary constraints that appear are the super-Hamiltonian and super-momentum, which in a general model can be considered as the simple sum of the contributions of the considered fields. So it can be shown that \emph{they preserve their role of generators of the diffeomorphisms and exhibit a closed algebra}.\\
This means that there are no tertiary constraints in the model, because the consistency conditions are fulfilled. The resulting Hamiltonian is vanishing as a consequence of the fact that the diffeomorphism invariance was never broken. 
At this point one is able to apply the Brown-Kucha\v r mechanism: squaring the super-momentum and imposing it on the super-Hamiltonian, one can solve the latter for $\pi$, obtaining something in the form $\pi - h = 0$. 
This way one gets
\begin{equation}\label{sgrhfunction}
\pi\pm\sqrt{\rho_0q\frac{d}{d-\Xi^2}}=0 = \pi  - h,
\end{equation}
where $\Xi = \sqrt{q} \rho_0 S T + H^G$ and $d = H^G_a H^G_b q^{ab}$. $\Xi$ is also the expression the super-Hamiltonian reduces to in the co-moving frame.\\
The $h$ function is now the candidate for the construction of the physical Hamiltonian, defined as in section \ref{thekbmechanism}:
\begin{equation}\label{sgrphyshamiltonian}
\mathcal{H}_{phys} = \int d^3 \! x \left(- \sqrt{\rho_0q\frac{d}{d-\Xi^2}}\right).
\end{equation}\\
It is a scalar density of weight one, as one can easily check remembering the weight of the functionals appearing in \eqref{sgrhfunction}, so its integral over the 3-dim manifold is invariant under the action of the 3-dim diffeomorphism group, as needed.\\
Moreover, $h$ does not contain the field $\phi$, so its Poisson brackets with $\pi$ are vanishing, 
as well as $\{\pi,\pi\}$, and the commutator $\{h(f), h(g)\}$ does not contain any dependence on $\pi$. 
Hence, $\{h(f), h(g)\}$ cannot reproduce any other constraint. Since the full system of constraints must 
be first-class ($H$ and $\pi - h$ are equivalent constraints) one can conclude, according with 
\cite{kucharbrown:1995}, that $\{h(f), h(g)\}$ strongly vanishes
\begin{equation} 
\{h(f), h(g)\}=0,
\end{equation}
for arbitrary smearing fields $f(x)$ and $g(x)$.

Finally, since the $h$ and $H_a$ generate respectively re-parametrization of the time-like variable and spatial diffeomorphisms, \eqref{sgrphyshamiltonian} is an observable.

As one can see the role of the entropy field $S$ is not very clear in \eqref{sgrphyshamiltonian}, but it enters explicitly in the time evolution. The general form of \eqref{sgrphyshamiltonian} is quite complicated and any attempt for quantization seems to be not viable. However the striking result is that the potential term that appears in the action gives to the entropy a privileged position in the definition of a time variable.\\
The direct interpretation of entropy as time variable comes out in the particular case of the co-moving frame. Again the conjugate momentum reduces to:
\begin{equation}
\pi = - \sqrt{q} \rho_0,
\end{equation}
while the set of secondary constraints becomes:
\begin{subequations}
\begin{align}
&\Xi = \sqrt{q} \rho_0 S T + H^G=0,\\
&\Xi_a = H^G_a=0.
\end{align}
\end{subequations}
One can use the super-Hamiltonian to obtain an expression for $\sqrt{q} \rho_0$, so that $\pi$ reads:
\begin{equation}
\pi = \frac{H^G}{ST}
\end{equation}

Recalling the expression for the momentum conjugate to $S$ \eqref{sconstraint} one can multiply both sides by $\theta$, so that:
\begin{equation}
S p_S = \frac{\theta H^G}{T} = \bar{h},
\end{equation}
which integrated over the spatial manifold as in \eqref{sgrphyshamiltonian} gives the equation:
\begin{equation}
\{ \mathcal{\bar{H}}_{phys}, \mathcal{O}_f(\tau) \} = \frac{d}{d ln S} \mathcal{O}_f(\tau).
\end{equation}
So \emph{one can identify the time parameter $\tau$ with the logarithm of the entropy per baryon}, and finally write:
\begin{equation}
-\frac{d}{d \tau} \mathcal{O}_f(\tau) = \{ \int d^3 \!x \frac{\theta H^G}{T}, \mathcal{O}_f(\tau) \},
\end{equation}
which is what one was searching for: a Schr\"odinger-like evolution equation for observables. 

\section{Concluding remarks}
A possible solution to the problem of time in canonical quantum gravity is proposed: using the Brown-Kucha\v r mechanism one is able to write an equivalent set of secondary constraints for GR coupled with the Schutz' perfect fluid. In this context the super-Hamiltonian acquires the form of a Schr\" odinger operator, i.e. becomes a physical Hamiltonian, which generates time evolution for observables.\\
It expression is complex, but the entropy field $S$ enters directly its definition: time and entropy are highly related.\\
This relation is much more evident in the co-moving frame, where the absence of the spatial gradients of the fluid allows one to find a clearer physical interpretation: in this particular case the logarithm of entropy per baryon is exactly the time variable.
Considering that the number of particles $n$ for the fluid is fixed \cite{schutz:1970}, one can simply obtain the entropy of the system by multiplying by $n$ the $S$ field. This is surprising, because there are no hints in the theory about that privileged role played by the entropy, and because $S$, and so its logarithm, are naturally future pointing quantities in a closed system.\\
Moreover the evolution operator is rather simple, above all because the matter-free super-Hamiltonian appears linearly in the physical Hamiltonian. The only fluid variable is $\theta$, whose equation of motion contains the arbitrary temperature function $T$, and whose conjugate momentum is vanishing. It is just an arbitrary field fixed by initial value conditions.

\section*{Aknowledgements}
We would like to thank M. Testa for his precious suggestions about the preservation of the diff-invariance in the coupled theory. 
The work of S.Z. is supported by the Belgian Federal Office for Scientific, Technical and Cultural Affairs through the Interuniversity Attraction Pole P6/11.

\bibliographystyle{ieeetr}

\end{document}